\documentclass[aps,prl,twocolumn,groupedaddress,nofootinbib,notitlepage,showpacs,floatfix]{revtex4-1}

\usepackage{graphicx}

\usepackage{color}
\usepackage{amsmath,amssymb}
\usepackage{rotating}
\usepackage[colorlinks, linkcolor = blue, citecolor = blue, filecolor = blue, urlcolor = blue]{hyperref}

\definecolor{orange}{rgb}{0.5,0.5,0}
\newcommand{\be}{\begin{equation}}
\newcommand{\ee}{\end{equation}}
\newcommand{\ba}{\begin{array}}
\newcommand{\ea}{\end{array}}
\newcommand{\bqa}{\begin{eqnarray}}
\newcommand{\eqa}{\end{eqnarray}}

\newcommand{\tr}{\mbox{Tr}}

\newcommand{\bra}[1]{\ensuremath{\langle #1 |}}
\newcommand{\ket}[1]{\ensuremath{| #1 \rangle}}

\begin{document}
\title{Completely positive approximate solutions of driven open quantum systems}

\author{Farhang Haddadfarshi}
\affiliation{Freiburg Institute for Advanced Studies, Albert-Ludwigs-Universit\"{a}t, Albertstra\ss e 19, 79104 Freiburg, Germany}
\affiliation{Department of Physics, Imperial College London, London SW7 2AZ, United Kingdom}
\author{Jian Cui}
\affiliation{Freiburg Institute for Advanced Studies, Albert-Ludwigs-Universit\"{a}t, Albertstra\ss e 19, 79104 Freiburg, Germany}
\affiliation{Department of Physics, Imperial College London, London SW7 2AZ, United Kingdom}
\author{Florian Mintert}
\affiliation{Freiburg Institute for Advanced Studies, Albert-Ludwigs-Universit\"{a}t, Albertstra\ss e 19, 79104 Freiburg, Germany}
\affiliation{Department of Physics, Imperial College London, London SW7 2AZ, United Kingdom}

\begin{abstract}
We define a perturbative approximation for the solution of Lindblad master equations with time-dependent generators that satisfies the fundamental property of complete positivity, as essential for quantum simulations and optimal control.
With explicit examples we show that ensuring this property substantially improves the accuracy of the perturbative approximation.
\end{abstract}

\maketitle

The properties of quantum systems can be modified in a seemingly limitless fashion through the application of external, coherent driving.
In closed systems, this can be used {\it e.g.} to generate entanglement among trapped ions \cite{ciraczoller,molmer}, or to realize effective Hamiltonians that mimic the physics of sophisticated solid state systems \cite{bloch,qpt1,sengstock}.
In open quantum systems, driving permits among others to modify interactions with a system's environment that make the system equilibrate at sub-Kelvin temperatures despite surrounding room-temperature \cite{philips,lasercooling},
or to stabilize many-body states with desired properties \cite{zoller2}.
Given this potential, the realization of quantum simulators that mimic the properties of different many-body systems has turned into one of the most actively pursued goals in research on quantum mechanical systems \cite{tobi,blatt-roos}.

A quantum simulator that helps us understand problems that we can not tackle with our conventional means, necessarily describes a system whose dynamics we can not solve.
Identifying the proper driving that helps to implement this dynamics, therefore, necessarily needs to be based on an analytic approximation, such as a perturbative treatment.
Most commonly, this implies a series expansion of the propagator \cite{PhysRevA.68.013820,PhysRevLett.111.175301,PhysRevX.4.031027},
like the {\em Magnus expansion} \cite{ME}, which approximates the propagator $U(t)=e^{-i\mathcal{K}(t)}$ through an expansion of its generator $\mathcal{K}(t)$.
Assuming periodic driving (with period $T$), the generator $\mathcal{K}(t)$ defines the effective Hamiltonian $\mathfrak{H}=\mathcal{K}(T)/T$, which satisfies $U(nT)=e^{-i\mathfrak{H}nT}$ for all integer $n$.
Under a stroboscopic perspective in which the system is monitored at multiples of driving periods only,
the effective Hamiltonian $\mathfrak{H}$ and $H(t)$ induce the same dynamics, so that a system with the time-dependent Hamiltonian $H(t)$ simulates the dynamics induced by $\mathfrak{H}$.

Quantum simulations are not necessarily limited to closed quantum systems,
and given the numerous big challenges in theory of open quantum systems,
an approach in terms of quantum simulations seems desirable \cite{nature1}.
Rather than unitarity, the fundamental properties of an equation of motion
$\dot\varrho={\cal L}(t)\varrho$
for the density matrix $\varrho$ of an open system with a time-dependent generator are complete positivity and preservation of trace.
Since $\tr{\cal L}(t)\varrho=0$, the Magnus expansion
$V_n(t)=e^{{\mathfrak{L}}_n(t)}$ with ${\frak L}_n(t)=\sum_{i=0}^n M_i(t)$ and
\begin{eqnarray}
M_0(t) &=&\int_0^tdt^{\prime}\mathcal{L}( t^{\prime}),\\
M_1(t) &=&\frac{1}{2}\int_0^tdt^{\prime}[\mathcal{L}(t^{\prime}),M_0(t^{\prime})] 
\end{eqnarray}
and similar higher order terms $M_i(t)$ \cite{ME} yields a trace-preserving approximation $V_n(t)$ of the exact propagator $V(t)$. An identification of ${\frak L}_n(T)/T$ with an effective generator of a to-be-simulated open system dynamics is, however, typically not possible, since ${\frak L}_n(T)/T$ does not necessarily induce completely positive dynamics.
Our goal is to modify the Magnus expansion in order to ensure that an approximate generator does indeed induce completely positive dynamics.

An essential feature of the Magnus expansion for {\em closed} systems is that it yields a unitary, approximate propagator in any order of approximation.
This is different in standard time-dependent perturbation theory in which the propagator itself is expanded in a series \cite{dyson}. Due to the exponential function, which is given by an {\em infinite} series,
the propagator with an $n$-th order generator contains terms of higher than $n$-th order, and it is exactly those terms that make sure that the fundamental property of unitarity is satisfied. In a similar fashion, we will strive for the incorporation of suitably chosen higher order terms to the generator that make sure that the fundamental property of complete positivity is satisfied in the treatment of open systems.

Unlike in the case of unitary dynamics, the characterization of generators of completely positive dynamics is a largely open question.
Only for infinitesimal dynamics does the celebrated Lindblad form \cite {lindblad}
\be
\mathcal{L}\circ=-i[H,\circ]+\sum_{ij}C_{ij}(\sigma_i\circ\sigma_j^\dagger-\frac{1}{2}\{\sigma_j^\dagger\sigma_i,\circ\})
\label{eq:lindblad}
\ee
with a positive-semidefinite, Hermitian coefficient matrix $C$ characterize all valid generators.
In general, however, Lindblad form is only a sufficient, but not a necessary condition for a valid generator \cite {wolf}.
Yet, as we will show with several explicit examples, an approximate generator can often be extended to be of Lindblad form consistently with the given order of expansion,
and comparison with the numerically exact solution demonstrates that the extended propagator provides a substantially better approximation than the original Magnus expansion.

The Magnus expansion (or any other series expansion \cite{PhysRevA.68.013820,PhysRevLett.111.175301,PhysRevX.4.031027}) at finite order $n$ provides an approximate generator $\mathfrak{L}_n$ of the form of Eq.~\eqref{eq:lindblad},
that permits to read off an effective Hamiltonian $\mathfrak{H}_n$ and an effective coefficient matrix $\mathfrak{C}_n$.
The central flaw is, that $\mathfrak{C}_n$ is not necessarily positive  semidefinite, so that complete positivity of the induced dynamics is not ensured.
$\mathfrak{C}_n$ is given in terms of the series $\mathfrak{C}_n=\sum_{i=0}^{n}\mathfrak{D}_i\omega^{-i}$, where $\omega=2\pi/T$ is the fundamental driving frequency.
Adding a term $\sum_{i=n+1}^{n^\prime}\mathfrak{D}_i\omega^{-i}$ with suitably chosen matrices $\mathfrak{D}_i$ ($i>n$) might result in a positive-semidefinite matrix
$\tilde{\mathfrak{C}}_n=\sum_{i=0}^{n^\prime}\mathfrak{D}_i\omega^{-i}$,
and the generator $\tilde{\mathfrak{L}}_n$
defined in terms of $\mathfrak{H}_n$ and $\tilde{\mathfrak{C}}_n$ would be a valid generator of open quantum system dynamics.

There is no unique choice for the matrices $\mathfrak{D}_i$ ($i>n$) which leaves substantial ambiguity in the construction of a valid generator,
but we found that the following modification of perturbatively obtained eigenvalues yields very good results:
typically it is possible to diagonalize $\mathfrak{C}_n$ in a perturbative manner with the expansion coefficient $1/\omega$.
The eigenvalues $\lambda_i$ are then approximated as $\lambda_i^{(n)}=\sum_{j=0}^n\mu_{ij}\omega^{-j}$.
As long as the leading contribution $\mu_{ij}$ of $\lambda_i^{(n)}$ is positive, one can always construct a $\tilde\lambda_i^{(n)}$,
such that $\lambda_i^{(n)}-\tilde\lambda_i^{(n)}$ is at least of order $\omega^{-(n+1)}$ and such that $\tilde\lambda_i^{(n)}$ is the square of a real quantity, and, correspondingly non-negative.
With the eigenstates $\Phi_i^n$ of $\mathfrak{C}_n$ in $n$-th order, one readily constructs the positive matrix
\be
\tilde{\mathfrak{C}}_n=\sum_i\tilde\lambda_i^{(n)} \Phi_i^{(n)} (\Phi_i^{(n)})^\dagger
\label{eq:corrected}
\ee
such that deviations between $\mathfrak{C}_n$ and $\tilde{\mathfrak{C}}_n$ are of higher than $n$-th order.

In the following we will discuss this scheme for some explicit examples of driven open quantum systems, and demonstrate that -- in addition to completely positive dynamics -- the corrected generators $\tilde{\mathfrak{L}}_n$ also provide a substantially better approximation of the exact dynamics than their counterpart $\mathfrak{L}_n$.

Let us begin with the driven two-level system whose Hamiltonian reads
\begin{equation}
H_{tl}(t)=\frac{\omega_0}{2}\sigma_z+(\Omega_s\sin(\omega t)+\Omega_c\cos(\omega t))\sigma_x\ ,
\nonumber
\end{equation}
in terms of the Pauli matrices $\{\sigma_i\}$,
the resonance frequency $\omega_0$ of the bare system,
and the driving amplitudes for {\it sine} and {\it cosine} driving.
In the presence of dephasing with rate $\gamma$\footnote{which itself may depend on the driving}, the time-dependent generator $\mathcal{L}_{tl}(t)$ reads
\be
\mathcal{L}_{tl}(t)\circ=-i[H_{tl}(t),\circ]+\gamma(\sigma_z\circ\sigma_z-\circ)\ . \nonumber
\ee
The effective Hamiltonian reads
\be
\mathfrak{H}_{tl}=\frac{\omega_0}{2}\sigma_z
+\frac{\omega_0\Omega_s}{\omega}\sigma_y+ 
\frac{1}{\omega^2}(A\sigma_x+ B \sigma_z )
+O(\omega^{-3})
\label{eq:effectiveH}
\ee
with $ A=-\omega_0/2(\Omega_c^2+3\Omega_s^2)$ and $B=(4\gamma^2-\omega_0^2)\Omega_c$.
The lowest and first order term coincide with the effective Hamiltonian of the coherent system \cite{PhysRevA.68.013820}, but in second order the effective Hamiltonian is actually influenced by the presence of dephasing.
The main impact of dephasing is, however, encoded in the coefficient matrices $\mathfrak{C}_n$ 
that are comprized (including third order) of
\begin{equation}
\ba{rclcrcl}
\mathfrak{D}_0&=&
\left[
\ba{ccc}
0&0&0\\
0&0&0\\
0&0&\gamma
\ea\right]\ ,
&
\mathfrak{D}_1&=&2\gamma\Omega_s
\left[
\ba{ccc}
0&0&0 \nonumber \\
0&0&1  \\
0&1&0
\ea\right]\vspace{.2cm}\ ,\\
\mathfrak{D}_2&=&
\left[
\ba{ccc}
0&0&\alpha \nonumber \\
0&\beta&0  \\
\alpha&0&-\beta
\ea\right]\ ,
&
\mathfrak{D}_3&=&
\left[
\ba{ccc}
0&\alpha^{\prime}&0\\
\alpha^{\prime}&0&\beta^{\prime}\\
0& \beta^{\prime} &0
\ea\right]\ ,  \nonumber
\ea
\end{equation}
with $\alpha=-4\gamma\omega_0\Omega_c$,
$\beta=2\gamma(\Omega_c^2+3\Omega_s^2)$, $\alpha^{\prime}=-6\gamma\omega_0\Omega_c\Omega_s$ and $\beta^{\prime}=-2\gamma\Omega_s(12\gamma^2-9\omega_0^2+12\Omega_c^2+20\Omega_s^2)$,
and the set of matrices $\{\sigma_i\}$ that help to define a Lindblad operator via Eq.~\eqref{eq:lindblad} is given by the Pauli matrices.

Since $\mathfrak{C}_0$ coincides with the time-independent coefficient matrix $C$, it is necessarily positive semidefinite and no correction is necessary. On the other hand
$\mathfrak{C}_1$ has an eigenvalue $-4\gamma\Omega_s^2/\omega^2$;
that is, $\mathfrak{C}_1$ is in general not positive semidefinite.
In first order approximation (including terms up to $\omega^{-1}$ only), however, all eigenvalues are non-negative.
We may therefore use $\tilde\lambda_1^{(1)}=\gamma$, $\tilde\lambda_2^{(1)}=\tilde\lambda_3^{(1)}=0$ and the perturbative eigenvectors
\be
\Phi_1^{(1)}=\left[\ba{c}0\\2\Omega_s/\omega\\1\ea\right],\
\Phi_2^{(1)}=\left[\ba{c}0\\1\\-2\Omega_s/\omega\ea\right],\
\Phi_3^{(1)}=\left[\ba{c}1\\0\\0\ea\right]\nonumber
\ee
to obtain the positive-semidefinite matrix
\be
\tilde{\mathfrak{C}}_1=\gamma\left[
\ba{ccc}
0&0&0\\
0&4\Omega_s^2/\omega^2&2\Omega_s/\omega\\
0&2\Omega_s/\omega&1
\ea\right]
\ee
using Eq.\eqref{eq:corrected},
which, together with Eq.\eqref{eq:effectiveH} defines the valid generator $\tilde{\mathfrak{L}}_1$.

The eigenvalues of $\mathfrak{C}_2$ including up to second order contribution read 
$\lambda_1^{(2)}=\gamma-2\gamma(\Omega_s^2+\Omega_c^2)/\omega^2$,
$\lambda_2^{(2)}=2\gamma(\Omega_s^2+\Omega_c^2)/\omega^2$ and
$\lambda_3^{(2)}=0$.
Since $\lambda_1^{(2)}$ can adopt negative values, it is necessary to introduce the non-negative modification 
\be
\tilde\lambda_1^{(2)}=\gamma\left(1-\frac{\Omega_s^2+\Omega_c^2}{\omega^2}\right)^2\ , \nonumber
\ee
whereas $\tilde\lambda_i^{(2)}$ coincides with $\lambda_i$ for $i=2,3$.
With the corresponding eigenvectors, one then obtains a positive- semidefinite matrix $\tilde{\mathfrak{C}}_2$ and corresponding generator $\tilde{\mathfrak{L}}_2$.
This scheme may readily be implemented also in higher order, but we will limit ourselves in the following to the discussion of up to third order.

\begin{figure*}[t]
\includegraphics[width=0.95\linewidth]{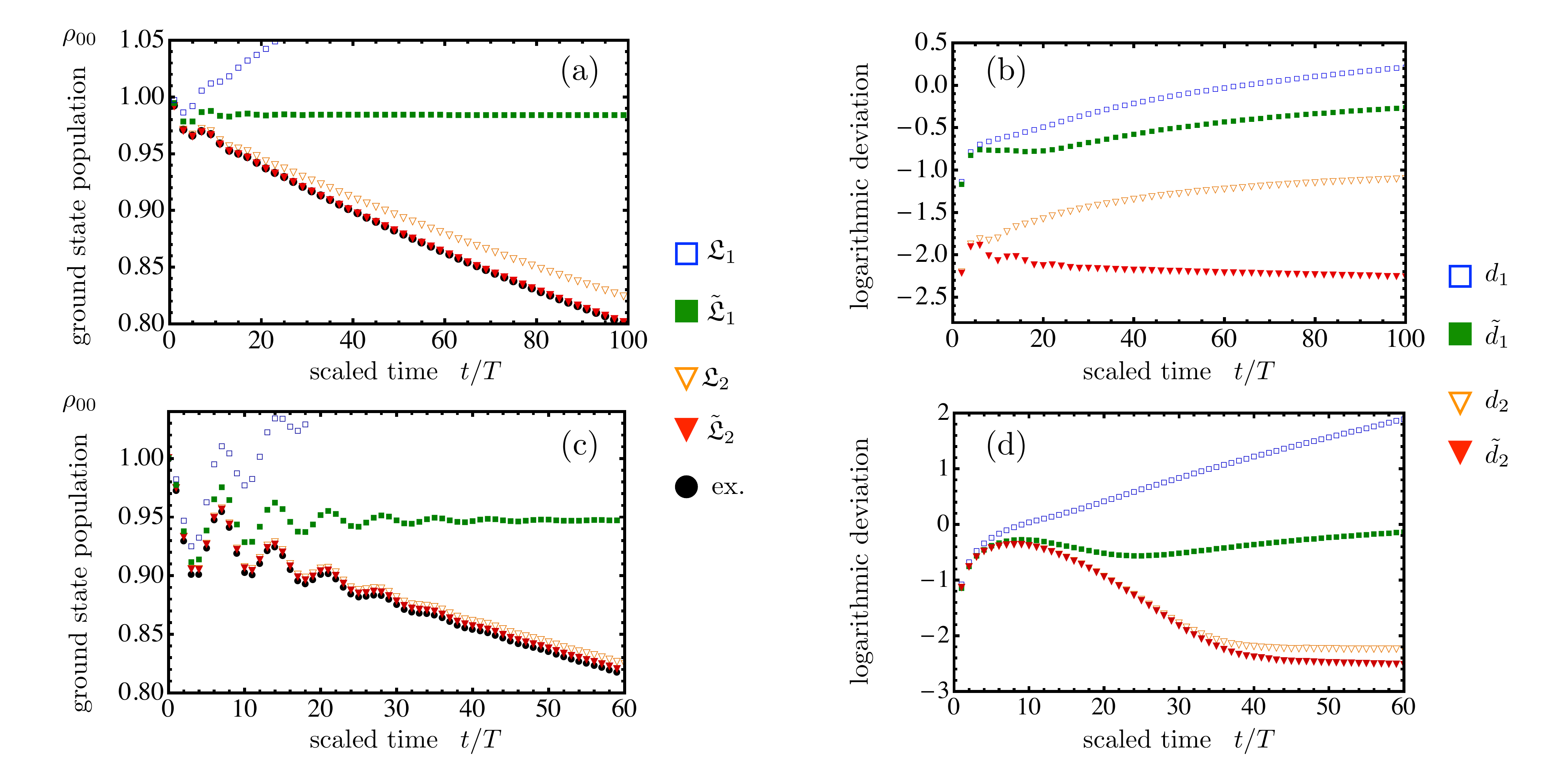}
\caption{(Color online)
Inset (a) depicts the ground state population of a driven, dissipative two-level system as function of time (in multiples of the driving period $T$) for the specific parameters $\Omega_s/\omega=1/10, \Omega_c/\omega=1/9$  and $\gamma/\Omega_s=1/8$.
Empty squares (blue) and triangles (orange) depict the Magnus approximation in first and second order, and full squares (green) and triangles (red) depict the corresponding completely positive approximations; the exact solution is depicted in full circles (black).
Inset (b) depicts the comparison defined in Eq.~\eqref{eq:distance} of the propagators in Magnus approximation and their extensions that demonstrates the added value of the present completely positive, approximate propagators.
Inset (c) and (d) show the analogue information for a driven harmonic oscillator with the parameters $\Omega/\omega=\gamma/\Omega=1/8$.}
\label{fig:1}
\end{figure*}

Fig.~\ref{fig:1} (a-b) depicts a comparison of the dynamics induced by the effective generators $\mathfrak {L}_n$,  in first and second order, their corrected counterpart $\tilde{\mathfrak{L}}_n$  and the exact effective dynamics for the specific parameters $\Omega_s/\omega=1/10, \Omega_c/\omega=1/9$  and $\gamma/\Omega_s=1/8$.
Inset (a) shows the population of the ground state $|0\rangle (=-\sigma_z|0\rangle)$ with $\varrho(0)=\ket{0}\bra{0}$ as initial condition.
After a short decline, the ground state population exceeds the value of $1$ in the dynamics induced by ${\frak L}_1$;
since the dynamics is trace-preserving, this implies a negative value of $\bra{1}\varrho\ket{1}$, so that the dynamics is clearly not completely positive.
The dynamics induced by $\tilde{\mathfrak{L}}_1$ overcomes this flaw, 
and provides a substantially better approximation of the exact dynamics; after a few driving periods, however, also $\tilde{\mathfrak{L}}_1$ ceases to approximate the exact dynamics well.
This is substantially improved with the second-order approximations, where both ${\frak L}_2$ and $\tilde{\frak L}_2$ yield a decent approximation in the entire depicted domain.
Similarly to the first order case, also $\tilde{\frak L}_2$ induces a more accurate approximation than ${\frak L}_2$,
and the same holds in third order (not shown).
Inset (b) underlines that this observation is independent of initial condition, but that the propagators $\tilde V_n$ induced by $\tilde{\frak L}_n$ are substantially more rigorous approximations to  the exact propagator than the propagators $V_n$ induced by $\frak L_n$.
Inset (b) depicts the logarithmic deviations
\be
d_n=\log||V(t)-V_n(t)||\mbox{ and }\tilde d_n=\log||V(t)-\tilde V_n(t)||
\label{eq:distance}
\ee
where $||\circ||^2=\tr(\circ\circ^{\dagger})$ denotes the Hilbert-Schmidt norm.
As one can see, a substantial deviation between $d_n$ and $\tilde d_n$ sets in after few driving cycles,
and the incorporation of completely positive dynamics results in an improvement of an order of magnitude,
and the improvement in third order (not shown) is of one order of magnitude as well.

This behaviour is by no means unique to the two-level system, but we found qualitatively the same behavior also for $\Lambda$-systems comprised of two degenerate ground-states and an excited state, or also for the harmonic oscillator
with Hamiltonian and Lindblad operator
\begin{eqnarray}
H_{ho}(t)&=&\omega_0\ a^{\dagger}a+\Omega\sin(\omega t)(a+a^{\dagger})\hspace{3.5 mm} \nonumber \\
\mathcal{L}_{ho}(t)\circ &=& -i[H_{ho}(t),\circ]+\gamma(\hat n\circ \hat n-\frac{1}{2}\{\hat n^2,\circ\})\nonumber
\end{eqnarray}
where $\{a , a^{\dagger}\}$ are the creation and annihilation operators with the commutation relation $[a,a^{\dagger}]=1$, $\hat n=a^{\dagger}a$ is the phonon-number operator and $\gamma$ is the dephasing rate.
Since the harmonic oscillator is an infinite dimensional system, the size of the coefficient matrix is not necessarily bounded,
but an expansion in finite order will always result in finite dimensional effective coefficient matrices $\mathfrak{C}_n$.

Including second order contributions, it is sufficient to use the operators $\{a,a^\dagger,\hat n\}$ as operator basis to define a generator via Eq.~\eqref{eq:lindblad} with the coefficient matrix
\begin{eqnarray}
\mathfrak{C}_2 &=& \begin{bmatrix}
0&0&0\\
0&0&0\\
0&0&\gamma
\end{bmatrix}+\frac{i\gamma\Omega}{\omega}\begin{bmatrix}
0&0&1\\
0&0&-1\\
-1&1&0\end{bmatrix}+\frac{3\gamma\Omega^2}{2\omega^2}\begin{bmatrix}
1&-1&0\\
-1&1&0\\
0&0&0
\end{bmatrix}
\nonumber
\end{eqnarray}
Neither $\mathfrak{C}_1$ nor $\mathfrak{C}_2$ are positive  semidefinite,
but the expansion of their eigenvalues consistent with the order of the underlying matrix yields non-negative quantities.
For $\mathfrak{C}_1$ one obtains $\lambda_1^{(1)}=\gamma$, $\lambda_2^{(1)}=\lambda_3^{(1)}=0$,
and for $\mathfrak{C}_2$ has $\lambda_1^{(2)}=\gamma+2\gamma\Omega^2/\omega^2$, $\lambda_2^{(2)}=\gamma\Omega^2/\omega^2$ and $\lambda_3^{(2)}=0$.
Consequently, the straight forward choice $\tilde\lambda_i^{(n)}=\lambda_i^{(n)}$ yields valid generators $\tilde{\mathfrak{L}}_n$ for $n=1,2$.
The insets (c) and (d) of Fig~\ref{fig:1} depict the performance of the obtained approximations for the specific parameters $\Omega/\omega=\gamma/\Omega=1/8$.
Inset (c) shows the ground state ($a|0\rangle=0$) population for the initial condition $\varrho(0)=|0\rangle\langle0|$.
Similarly to inset (a), $\mathfrak{L}_1$ induces dynamics that is evidently not compatible with the probabilistic interpretation of quantum mechanics which is salvaged by $\tilde{\mathfrak{L}}_1$,
and $\tilde{\mathfrak{L}}_2$ induces a much more accurate approximation of the exact dynamics than $\mathfrak{L}_2$. 
Similarly to inset (b), inset (d) demonstrates that substantial improvement in accuracy is obtained through the use of corrected generators\footnote{In the case of the harmonic oscillator Eq.~\eqref{eq:distance} is evaluated in the subspace spanned by the lowest four oscillator eigenstates}.
There is, however, a substantial difference between the harmonic oscillator and the two-level system:
the coefficient matrix $\mathfrak{C}_3$ \footnote{
$\mathfrak{C}_3$ is a $9\times 9$ matrix that defines a generator via Eq.\eqref{eq:lindblad} with the operator basis $\{a,a^\dagger,\hat n, a^{\dagger}a^2, {a^{\dagger}}^2a, {a^{\dagger}}^2a^2, {a^{\dagger}}^2a^3 , {a^{\dagger}}^3a^2, {a^{\dagger}}^3a^3 \}$}
has an eigenvalue with a negative third order term $\lambda_n^{(3)}=-3\sqrt{2}\gamma^3\Omega/\omega^3$.
Since this is the leading order contribution, it is not possible to find suitable higher order terms that help to define a valid generator.
Fundamentally, this is an issue that can be resolved only with the characterization of all valid generators of finite completely positive maps, but the present examples suggest that in practice a construction at sufficiently low order yields sufficiently good results that such cases can be avoided.

\begin{figure}[t]
\includegraphics[scale=0.185]{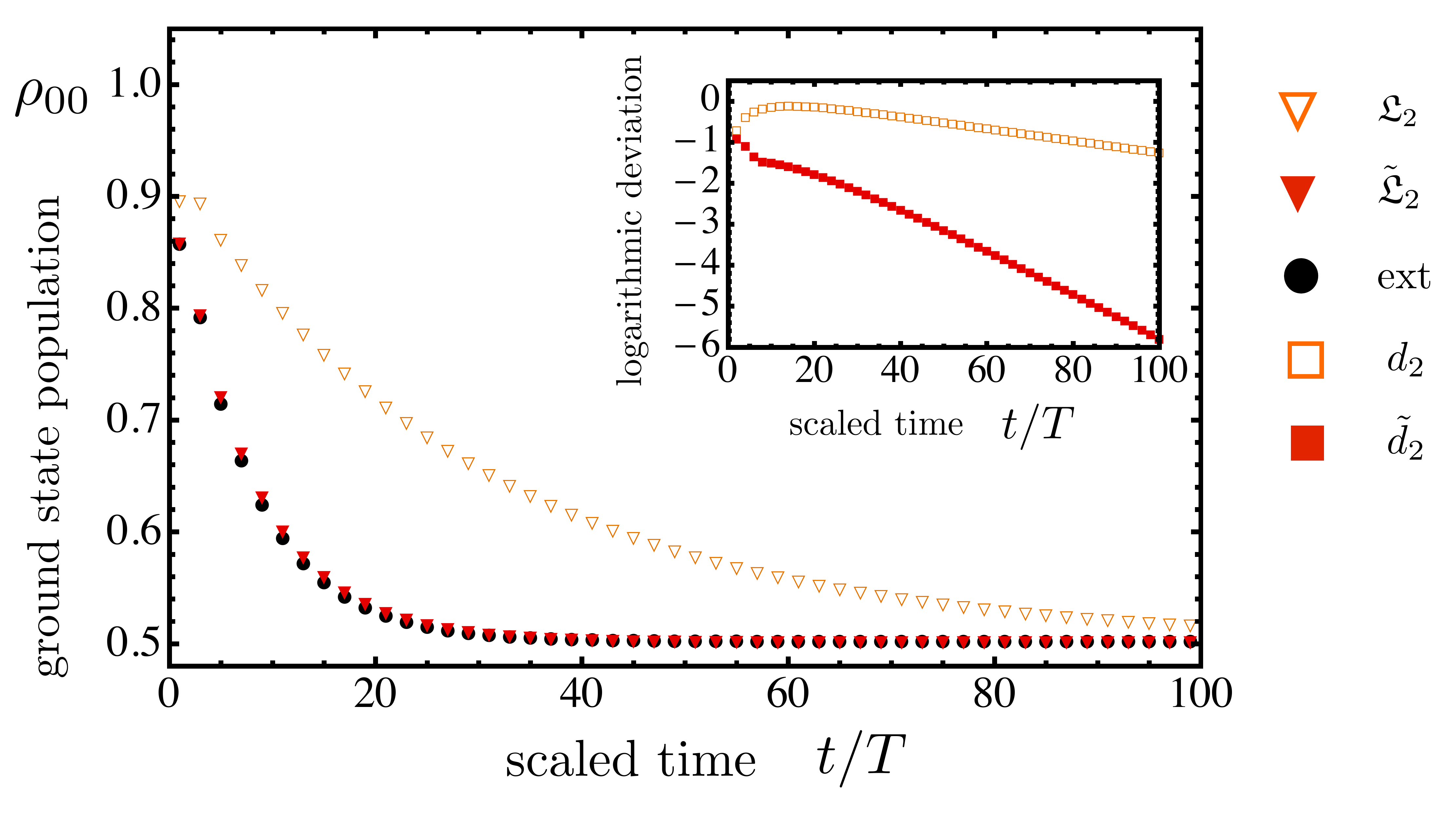}
\caption{Ground state population for a strongly driven dissipative two-level system with the parameters
$\Omega/\omega=\gamma/\Omega=1/3$.
Despite the failure of the second order Magnus approximation (empty triangle (orange)) to approximate the exact dynamics (circle (black)), the corrected Magnus expansion in second order (full triangle (red)) induced by $\tilde{\mathfrak{L}}_2$ provides an excellent approximation as also indicated by the logarithmic deviation (Eq.\eqref{eq:distance}) depicted in the inset.}
\label{fig:2}
\end{figure}

The term `{\em sufficiently low order}' implies applicability in the regime of weak and/or fast driving only where the Magnus expansion in low order is a good approximation.
We would, however, like to point out that the use of corrected generators $\tilde{\mathfrak{L}}_n$ also helps to expand the range of applicability substantially 
as one may see in Fig.~\ref{fig:2}; it depicts the ground state population of a driven two-level system in the extreme regime of strong driving with the parameters $\Omega/\omega=\gamma/\Omega=1/3$.
Even in second order, the regular Magnus expansion fails to approximate the actual dynamics, as it is also indicated by the inset that depicts the logarithmic deviation between the actual dynamics an the second order Magnus approximation according to Eq.~\eqref{eq:distance}.
The corrected generator $\tilde{\mathfrak{L}}_2$ on the other hand induces a highly accurate approximation,
what underlines the added value of constructing generators that induce valid (completely positive) dynamics.
This is of particular importance especially for variational analyses  aiming at the maximization of a specific goal like the preparation of a desired state or the implementation of a quantum gate.
If the variation of a control parameter that increases the value of the figure of merit (target functional) at the same time results in a reduced accuracy of the employed approximation, the seemingly optimal control parameters might induce anything but the desired dynamics, but a theory that ensures that all utilized  approximations satisfy a system's fundamental properties will prevent that.
The framework derived here is thus ideally suited for the design of optimal implementations of quantum simulations of open systems.

We are indebted to fruitful discussion with Albert Verdeny Vilalta, \L ukasz Rudnicki and Robert Alicki.
Financial support by the European Research Council within the project odycquent is gratefully acknowledged.

\bibliography{ref.bib}

\end{document}